\documentclass[prb,onecolumn,showpacs,superscriptaddress]{revtex4-2}

\usepackage{rotating,graphicx}
\usepackage{setspace}
\usepackage{amsmath}
\usepackage{mathtools}
\usepackage{tabularx}
\usepackage{mathrsfs}
\usepackage{lipsum}
\usepackage{xcolor}
\usepackage[citecolor=blue,linkcolor=blue,colorlinks=true,linkbordercolor=white]{hyperref}
\usepackage{float}
\usepackage{tikz}
\usepackage{pgfplots}
\usepackage{pifont}
\pgfplotsset{compat=newest}

\usepackage{braket}

\newcommand{\be}{\begin{equation}}
\newcommand{\ee}{\end{equation}}
\newcommand{\bea}{\begin{eqnarray}}
\newcommand{\eea}{\end{eqnarray}}

\newcommand{\la}{\langle}
\newcommand{\ra}{\rangle}
\newcommand{\ua}{\uparrow}
\newcommand{\da}{\downarrow}

\definecolor{azul}{HTML}{1F77B4}
\definecolor{laranja}{HTML}{FF7F0E}
\definecolor{verde}{HTML}{2CA02C}
\definecolor{vermelho}{HTML}{D62728}
\definecolor{roxo}{HTML}{9467BD}
\definecolor{marrom}{HTML}{8C564B}
\definecolor{rosa}{HTML}{E377C2}
\definecolor{cinza}{HTML}{7F7F7F}
\definecolor{amarelo}{HTML}{BCBD22}

\begin{document}

\title{Effects of Temperature and Magnetization on the Mott-Anderson Physics in one-dimensional Disordered Systems}

\author{G. A. Canella}
\affiliation{Institute of Chemistry, São Paulo State University, 14800-090, Araraquara, São Paulo, Brazil}

\author{K. Zawadzki}
\affiliation{Department of Physics, Royal Holloway University of London, TW10 0EX, Egham, UK}
\affiliation{ICTP South American Institute for Fundamental Research, 01140-070, São Paulo, Brazil}

\author{V. V. França}

\affiliation{Institute of Chemistry, São Paulo State University, 14800-090, Araraquara, São Paulo, Brazil}

\begin{abstract}

We investigate the Mott-Anderson physics in interacting disordered one-dimensional chains through the average single-site entanglement quantified by the linear entropy, which is obtained via density-functional theory calculations. We show that the minimum disorder strength required to the so-called full Anderson localization $-$ characterized by the real-space localization of pairs $-$ is strongly dependent on the interaction regime. The degree of localization is found to be intrinsically related to the interplay between the correlations and the disorder potential. In magnetized systems, the minimum entanglement characteristic of the full Anderson localization is split into two, one for each of the spin species. We show that although all types of localization eventually disappear with increasing temperature,  the full Anderson localization persists for higher temperatures than the Mott-like localization.

\end{abstract}

\maketitle

\section{Introduction}

The metal-to-insulator transition (MIT) in a nanostructure can be induced by the Coulomb interaction, as proposed by Mott-Hubbard \cite{16ref1,17ref1} or by disorder, as proposed by Anderson \cite{14ref1}. In the presence of both correlations and randomness one faces the interesting and far from be fully understood Mott-Anderson physics \cite{1, 2, 5, 7, 8, 10, 21, 22, ent1, ent2, francaAmico2011}.

Theoretical investigations of MIT in complex systems via exact methods are challenging and restricted to small systems. Must of the studies applies instead dynamical mean-field theory (DMFT) \cite{1ref6}, which properly accounts for the electronic interaction and the disorder potential, but are still demanding and limited to simple systems. 

Recently we have proposed an alternative approach in which the quantum entanglement $-$ quantified via density-functional theory (DFT) \cite{6ref6, 7ref6} calculations $-$ is used to explore the MIT in interacting disordered chains \cite{prb}. This methodology has been proven to be reliable when compared to exact density-matrix renormalization group (DMRG) data and has been also successfully applied to investigate the superfluid-to-insulator transition (SIT) in disordered superfluids \cite{gui1, gui2}. In both MIT and SIT cases entanglement was found to be a witness of {\it i)} the so-called full Anderson localization, associated to a real-space localization of pairs; {\it ii)} the Mott localization and {\it iii)} the Mott-like localization, associated to an effective density phenomenon. However the MIT study \cite{prb} was restricted to a fixed interaction strength, non-magnetized systems and at zero temperature.

Here we apply the same methodology to explore the Mott-Anderson physics in all the regimes of interaction and to investigate the impact of the magnetization and of the temperature in the MIT. We find that the minimum disorder strength necessary to the full Anderson localization is strongly dependent on the interaction regime. We also find an intrinsic connection between the level of the localization and the interplay between interaction and disorder. In magnetized systems, we find that the minimum entanglement characterizing the full Anderson localization is split into two minima, one for each spin species. Although the temperature fades away all types of localization, our results reveal that the full Anderson localization survives for higher temperatures than the Mott-like localization.

\section{Theoretical Model}

We simulate the disordered interacting lattices via the one-dimensional Hubbard model, 

\be{}
H = -t\sum_{\la ij \ra \sigma}(\hat{c}^{\dagger}_{i\sigma}\hat{c}_{j\sigma}+ H.c.) + 
U\sum_i\hat{n}_{i\ua}\hat{n}_{i\da} + \sum_{i\sigma}V_i\hat{n}_{i\sigma},
\ee{}
\noindent with on-site disorder potential $V_i$ characterized by a certain concentration $C\equiv L_V/L$ of randomly distributed impurities, where $L_V$ is the number of impurity sites and $L$ the chain size. The density operator is $\hat{n}_{i\sigma} = \hat{c}^{\dagger}_{i\sigma}\hat{c}_{i\sigma}$, the average density is $n=N/L=n_\uparrow+n_\downarrow$ and the magnetization is $m=n_\uparrow-n_\downarrow$, where $N = N_{\ua} + N_{\da}$ is the total number of particles and $\hat{c}^{\dagger}_{i\sigma}$ ($\hat{c}_{i\sigma}$) is the creation (annihilation) operator, with $z$-spin component $\sigma = \ua,\da$ at site $i$. All the energies are in units of $t$ and we set $t=1$.

We consider the average single-site entanglement: a bipartite entanglement between each site with respect to the remaining $L-1$ sites averaged over the sites. This ground-state entanglement is quantified via the average linear entropy,

\begin{equation}
\mathcal L=\frac{1}{L}\sum_i \mathcal L_i=1-\frac{1}{L}\sum_i\left(\text{w}_{\uparrow,i}^2+\text{w}_{\downarrow,i}^2+\text{w}_{2,i}^2+\text{w}_{0,i}^2\right),
\end{equation}
where $w_{\uparrow,i}$,$w_{\downarrow,i}$, $w_{2,i}$ e $w_{0,i}$ are the occupation probabilities for the four possible states of site $i$: single occupation with spin up, single occupation with spin down, double occupation and empty, respectively. At finite temperature,  the probabilities are calculated with respect to a thermal state $\rho_\beta = \sum_n e^{-\beta E_n} \ket{n} \bra{n}$, where $\ket{n}$ is an eigenstate of the Hamiltonian with energy $E_n$ and $\beta = 1/k_B T$ is the inverse temperature.
Thus for small chains ($L=8$) we calculate Eq.(2) by diagonalizing the full Hamiltonian.

We also explore larger ($L=100$) disordered chains at $T=0$ via density-functional theory calculations. In this case, instead of Eq.(2), we adopt an approximate density functional \cite{francaAmico2011} for the linear entropy of homogeneous chains,
\begin{equation}
\mathcal L^{hom}(n,U>0)\approx2n-\frac{3n^2}{2}+[(4n-2)\alpha(U)-4\alpha(U)^2]\times\Theta[n-\alpha(U)-1/2],
\end{equation}
where $\Theta(x)$ is a step function, with $\Theta(x)=0$ for $x<0$ and $\Theta(x)=1$ for $x\geq 0$, and $\alpha(U)$ is given by
\begin{equation}
\alpha(U)=2\int_0^\infty \frac{J_0(x)J_1(x)\exp\left[Ux/2\right]}{\left(1+\exp[Ux/2]\right)^2},
\end{equation}
where $J_k(x)$ are Bessel functions of order $k$. This density functional, Eq.(3), was specially designed to be used in LDA approximations for calculating the linear entropy of inhomogeneous systems via DFT calculations. Thus the entanglement in our large disordered chains are approximately obtained via LDA:
  
  \begin{equation}
\mathcal L\approx \mathcal L^{LDA}\equiv \frac{1}{L}\sum_i\mathcal L^{hom}(n,U>0)|_{n\rightarrow n_i},
\end{equation}
where the density profile $\{n_i\}$ is calculated via standard (Kohm-Sham iterative scheme) DFT calculations within LDA for the energy, in which the exact Lieb-Wu \cite{lw} energy is used as the homogeneous input.

For each set of parameters ($C,V;U,n,m$), $\mathcal L$ is obtained through an average over $100$ samples of random disorder samples to ensure that the results are not dependent on specific configurations of impurities. Notice that this huge amount of data would be impracticable via exact methods such as DMRG (for a comparison between our DFT approach and DMRG calculations, see Supplementary Material of Ref.\cite{prb}). 

\begin{figure} [t!]
    \centering
    \includegraphics[width=\textwidth]{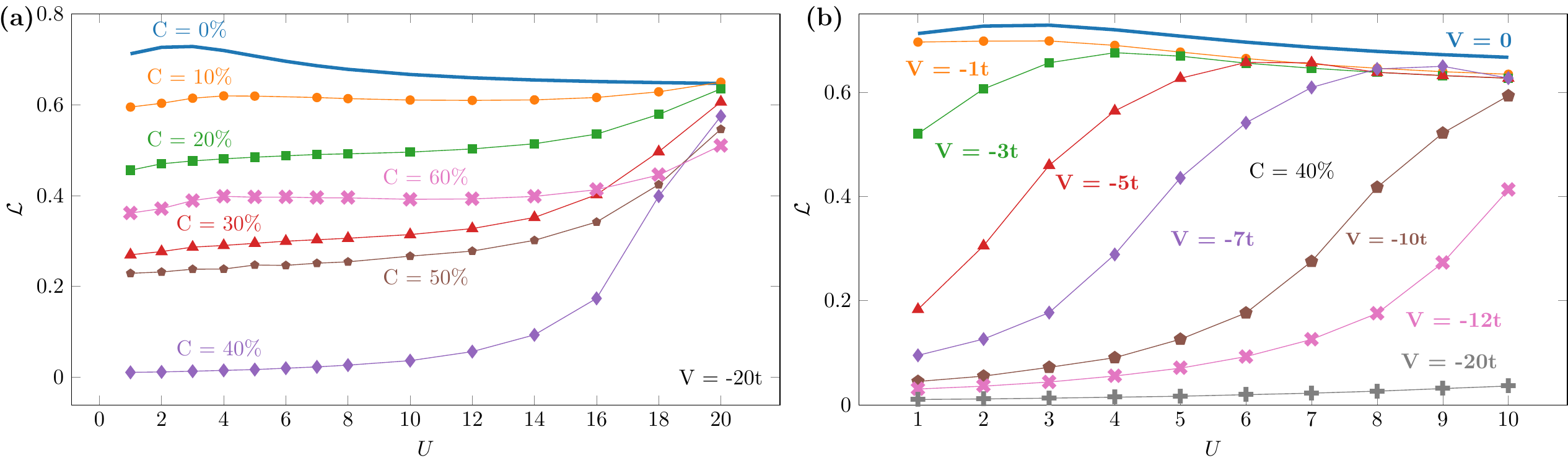}
\caption{Entanglement of disordered nanostructures as a function of the particle interaction: (a) for several concentrations $C$ of impurities with strength $V=-20t$ and (b) for several disorder strengths $V$ at the critical concentration $C_C=100n/2=40\%$.}
\end{figure}

\section{Results and Discussion}

We start by exploring the Mott-Anderson physics at zero temperature via the entanglement as a function of interaction. In Figure 1a we consider several concentrations of impurities with a fixed strength $V=-20t$, thus ranging from strong ($|V|>>U$) to moderate ($|V|\approx U$) disorder. As disorder becomes more relevant, i.e. for $U\rightarrow 0$, entanglement decreases and saturates for any concentration $C>0$. This saturation characterizes the localization: the disorder potential freezes the electronic degrees of freedom such that $\mathcal L\rightarrow \text{constant}$.

Fig. 1a also shows the non-monotonic behavior of entanglement with $C$, whose minimum occurs at the critical concentration $C_C=100n/2=40\%$ for $V<0$ (for $V>0$, $C_C=100(1-n/2)$), observed previously in the MIT \cite{prb} and in the SIT \cite{gui1,gui2}. This minimum entanglement has been associated $-$ in both MIT and SIT cases $-$ to a {\it fully localized} state, marked by $\mathcal L\rightarrow 0$ for $|V|\rightarrow \infty$ due to real-space localization of pairs (as also confirmed by the average occupation probabilities, see Figure 2). While for the MIT the full localization was found to appear for $|V|\geq V_{min}\approx 3t$ for $U=5t$, for the SIT the same minimum $V_{min}\approx 3t$ was found for any interaction \cite{gui2}. However we observe now a distinct feature: Fig. 1a reveals that depending on the interaction strength ($U>10t$) even a strong disorder potential as $V=-20t$ is not enough to fully localize the system,  i. e. $\mathcal L\neq 0$ at $C_C=40\%$. In other words, $V_{min}$ in the MIT case is strongly affected by the interaction.

To further analyze this interplay between $U$ and $V$, in Fig. 1b we focus on the critical concentration $C_C$ and vary instead the potential strength $V$. We find that for $|V|\lesssim U$ the degree of entanglement is essentially independent on $V$, suggesting that the system presents the same degree of localization for a given $U$ and that this is a weak localization, since the degree of entanglement is very close to the clean $V=0$ case. In contrast, for $|V|\gtrsim U$ the degree of entanglement decreases with $U$ decreasing and is smaller for higher $|V|$, reaching the full localization when $|V|\rightarrow \infty$. As $U$ increases, a stronger $V$ is required for having $\mathcal L \rightarrow 0$, confirming thus that the full localization in the MIT requires a minimum disorder strength $V_{min}$ which is dependent on the interaction. For the particular case of $V=-1t$, such that $U$ is always $U\gtrsim |V|$, we don't find the characteristic decreasing of entanglement when $U\rightarrow 0$, indicating that the full localization does not occur in this case. 

\begin{figure} [t!]
    \centering
    \includegraphics[scale=.65]{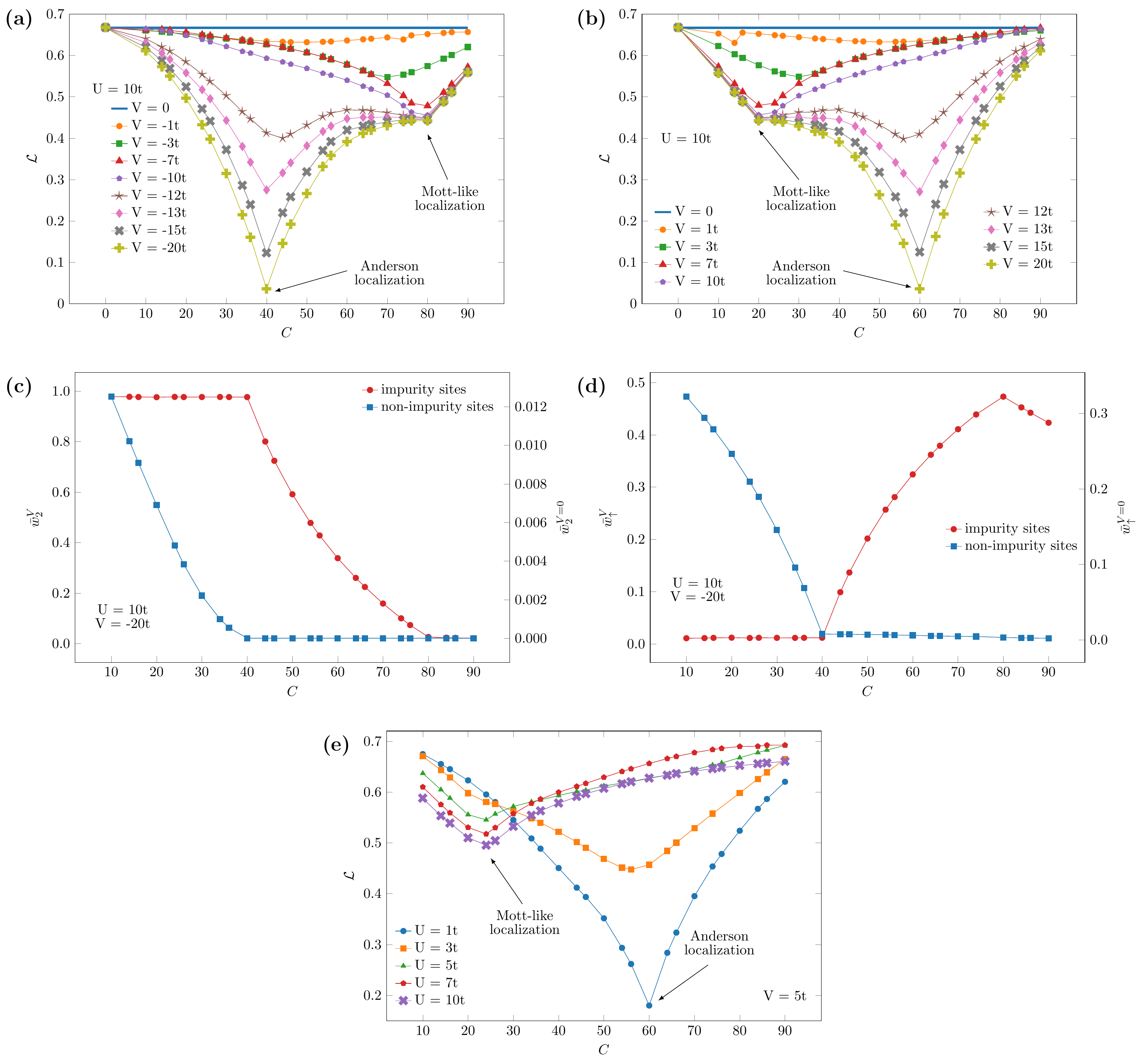}
    
    \caption{Entanglement of disordered nanostructures as a function of the impurities' concentration for several attractive (a) and repulsive (b) disorder strengths at a fixed $U$, and for several interaction strengths at a fixed $V$ (c). (d) and (e) average occupation probabilities as a function of the impurities concentration: double occupancies (d) and single-occupation probabilities (e) at impurity ($\bar{\text{w}}_2^V$, $\bar{\text{w}}_\uparrow^V$) and non-impurity sites ($\bar{\text{w}}_2^{V = 0}$, $\bar{\text{w}}_\uparrow^{V = 0}$). In all cases $L=100$ and $n = 0.8$.}
  
\end{figure}

\begin{figure}[t!]
    \centering
    \includegraphics[scale=.9]{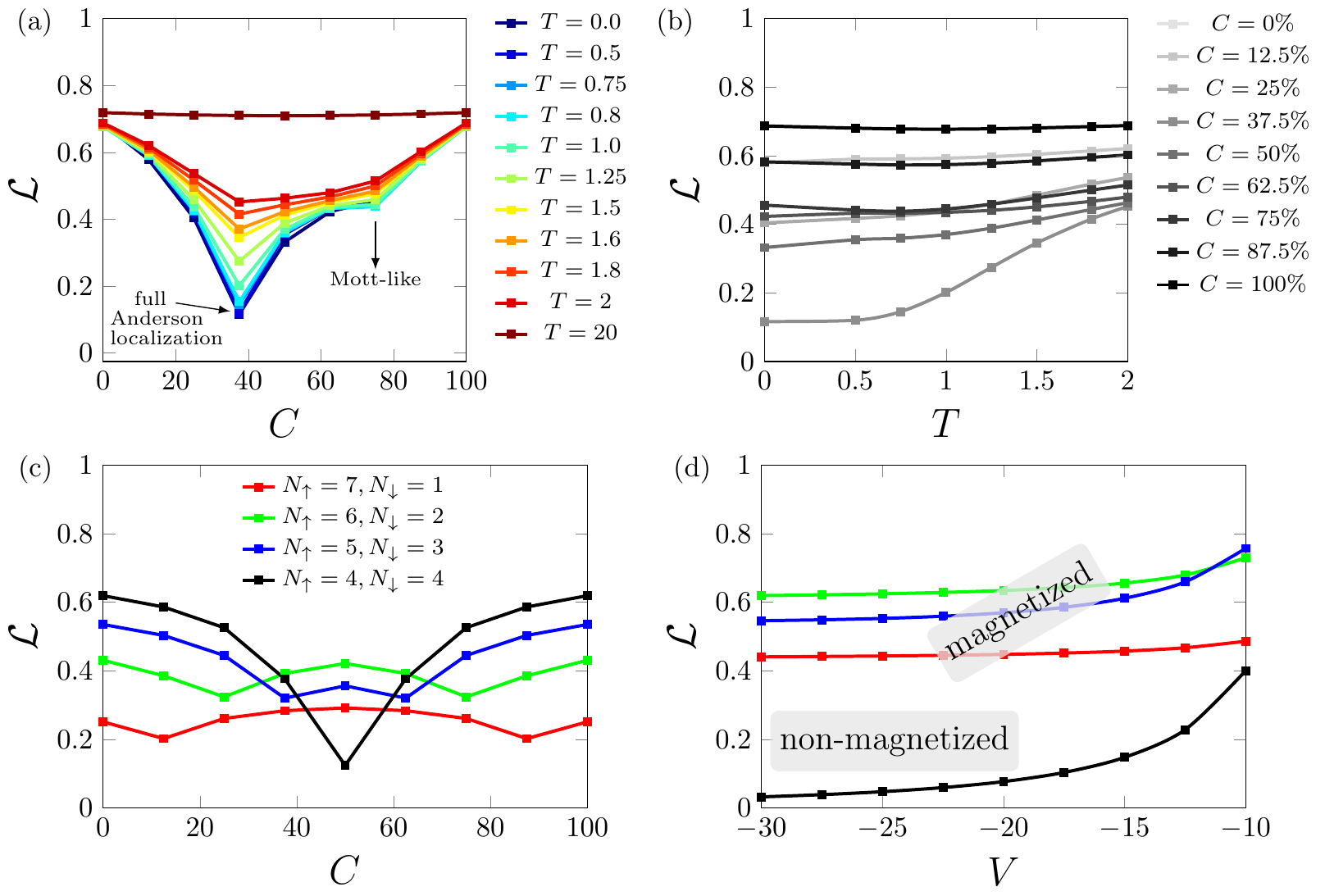}
    \caption{(a) Entanglement as a function of the concentration of impurities for several temperatures for $n=0.75$. (b) Entanglement as a function of the temperature for several concentrations for $n=0.75$. (c) Entanglement as a function of concentration for several magnetizations $m=n_\uparrow - n_\downarrow$ for $n=1.0$. In all cases $L=8$, $U=5t$ and $V=-10t$.}
\end{figure}

Next we analyze the impact of the impurities' concentration on the entanglement for several attractive, Figure 2a, and repulsive, Figure 2b, disorder strengths. In both cases we see the signature of the full Anderson localization for $|V|\gtrsim U$: minimum entanglement at the critical concentration $C_C=100n/2$ for $V<0$ and $C_C=(1-n/2)100$ for $V>0$, with $\mathcal L\rightarrow 0$ for $|V|\rightarrow \infty$. For $|V|\lesssim U$ the minimum at $C_C$ disappears, so the system does not fully localize. 

We also see the extra minimum at $C^\text{\small *}_C=100n$ for $V<0$ (Fig. 2a) and at $C_C^\text{\small *}=(1-n)100$ for $V>0$ (Fig. 2b) associated to a Mott-like localization \cite{prb}, in which the effective density is equal to 1 either at the impurity sites (for $V<0$) or at the non-impurity sites (for $V>0$). For attractive disorder this means that the average double occupancy in the impurity sites ($\bar{\text{w}}_2^V$) tends to zero due to the repulsion $U$, while the single-particle probability ($\bar{\text{w}}_\uparrow^V$) tends to a maximum, as confirmed by Figures 2c and 2d (for repulsive disorder, the same holds for the non-impurity sites: $\bar{\text{w}}_2^{V=0}\rightarrow 0$, $\bar{\text{w}}_\uparrow^{V=0}\rightarrow \text{maximum}$). Notice however that the Mott-like MIT requires a minimum amount of disorder to occur. Thus the two entanglement minima $-$ full Anderson and Mott-like localizations $-$ are intrinsically connected through the interplay between interaction and disorder. In Figure 2e one can see that if the interaction is too small compared to the disorder strength ($U\lesssim |V|/2$) only the minimum related to the full Anderson localization persists, while if $U$ is strong compared to $V$ ($U\gtrsim |V|$) only the minimum related to the Mott-like localization holds, the two minima appearing only for $U\gtrsim 10t$, $|V|\gtrsim U$.

In Figures 3a and 3b we show the impact of the temperature on both the full Anderson and the Mott-like localization. As the temperature increases the two minima $-$ at $C=100n/2=37.5\%$ (full Anderson) and at $C=100n=75\%$ (Mott-like) $-$ are attenuated. Our results reveal that the full Anderson localization survives for higher temperatures than the Mott-like localization, however for $T=20$ there remains no localization in the system, since entanglement is high and maximum for any concentration.

Finally, while all the above calculations were performed with non-magnetized chains, i. e. for $n_\uparrow=n_\downarrow=n/2$, in Figure 3c we analyze the impact of the magnetization $m=n_\uparrow - n_\downarrow\neq 0$ on the entanglement minimum related to the full Anderson localization. We find that the minimum at $C_C=100n/2=50\%$ for $m=0$ is now split into two minima: one at $C_C=100n_\uparrow$ and the other at $C_C=100n_\downarrow$. Our results thus reveal that the localization occurs separately for each species, thus with two critical densities $n_{C,\uparrow}=L_V$ and $n_{C,\downarrow}=L_V$. Fig.3d shows however that the magnetized systems never reach the full localization: there remain spin degrees of freedom due to the unpaired majority species such that $\mathcal{L}$ saturates finite values.

\section{Conclusion}

In summary, we have explored the Mott-Anderson physics by analyzing the entanglement of interacting disordered chains. We find that the interplay between interaction (U) and disorder strength (V) defines the type and the degree of localization. For weak interactions, $U\lesssim |V|/2$, there only appears the full Anderson localization, marked by entanglement approaching zero when $|V|\rightarrow \infty$. In contrast, for weak disorder, $|V|\lesssim U$, only the Mott-like localization holds, associated to an effective density equal to 1. The two types of localization, full Anderson and Mott-like, occurring only when both $U$ and $V$ are strong enough: $U\gtrsim 10t$, $|V|\gtrsim U$. For sufficiently strong interaction, $U\gtrsim |V|$, the entanglement is independent on the disorder potential and very close to the clean (non-disordered) case, suggesting thus that the localization is weak in this case. Our results also show that the temperature fades the localization phenomena, but that the full Anderson localization minimum survives for higher temperatures ($T\sim 2$) than the Mott-like localization.  Finally we have shown that 
that the entanglement minimum related to the full Anderson localization is split into two when there is a magnetization in the system, one for each spin species, but in this case the localization is weaker due to remaining spin degrees of freedom, with entanglement saturating at finite values.

\section{Data Availability}
The datasets generated during and/or analysed during the current study are available from the corresponding author on reasonable request.

\bibliography{references_MITAentanglement.bib}

\begin{center}\bf ACKNOWLEDGMENTS \end{center}

GAC was supported by the Coordena\c{c}ão de Aperfei\c{c}oamento de Pessoal de Nivel Superior - Brasil (CAPES) -
Finance Code 001. KZ was supported by FAPESP (Grants No 2016/01343-7 and 2020/13115-4). VVF was supported by FAPESP (Grants No 2019/15560-8 and 2021/06744-8). 

\section{Author Contributions}
G.A.C. and K.Z. performed the calculations. G.A.C., K.Z. and V.V.F. discussed the results and contributed to the writing of the manuscript.
\section{Additional Information}
Competing Interests: The authors declare no competing interests.


\end{document}